\documentclass[journal]{IEEEtran}

\usepackage[cmex10]{amsmath}

\newtheorem{proposition}{Proposition}
\newtheorem{definition}{Definition}
\newtheorem{theorem}{Theorem}

\def\QR#1#2{b^{(#2)}_{#1}}

\begin{document}

\title{Information Theory and Quadrature Rules}

\author{James~S.~Wolper
\thanks{James~S.~Wolper is with the Department of Mathematics,
Idaho State University, Pocatello, ID~83209-8085 USA.
email: wolpjame@isu.edu}}

\maketitle

\begin{abstract}
Quadrature rules estimate  $\int_0^1 f(x)~dx$ when $f$ is defined by
a table of $n+1$ values.  Every binary string of length $n$ defines
a quadrature rule by choosing which endpoint of each interval represents
the interval.  The standard rules, such as Simpson's Rule, correspond
to strings of low Kolmogorov complexity, making
it possible to define new quadrature rules with no
smoothness assumptions, as well as in higher 
dimensions.  Error results depend on concepts from compressed sensing.
Good quadrature rules exist for ``sparse" functions,
which also satisfy an error--information duality principle.  
\end{abstract}

\IEEEpeerreviewmaketitle

\section{Introduction}

\IEEEPARstart{R}{esearchers} have been showing how information
theory clarifies results about mathematics
and computing ever since Shannon \cite{shannon} defined the basic 
concepts.  This work considers {\sl quadrature\/} or
{\sl numerical integration\/} from an information theory perspective. 
The basic problem is to estimate $\int_0^1 f(x)~dx$ from a table
of values $f_i = f({i\over n})$, $i = 0, \ldots, n$.  This kind of problem 
arises naturally in applications, where, for example, one may only be able to
estimate the value of a function during a satellite pass, or at a
discrete set of ambient conditions such as temperature, or, in the social sciences, 
on Tuesdays.

Standard works on numerical analysis ({\it eg\/}, \cite{burden}, \cite{hamming}) 
develop quadrature methods that require one of two conditions that
are impossible to guarantee.  Many methods ({\it eg\/}, Gaussian
quadrature) require evaluation of $f$ at arbitrary points in its 
domain, which is impossible in the situation at hand.  Other methods
({\it eg\/} Newton--Cotes integration; see below) impose smoothness
conditions on $f$.  This, too, is problematic: imagine the effect of earthquake,
phase transition, or scandal on the functions whose measurement is
described above.

Estimation without control over the error is unsatisfying.
Integrating a function from a table is a kind of signal {\sl processing\/},
and ideas from signal {\sl reconstruction\/} lead to two error estimates,
at least for functions that have a sparse (although perhaps unknown)
representation.  The first is a kind of {\sl error--information duality\/}
for integration; briefly, the information in the error is the error in the
information. The second
is the existence of good quadrature rules for sparse functions.
Section V has details, including the definition of sparse.  

Here is the outline.   Section II  defines a {\sl primitive quadrature rule\/}
for estimating $\int f(x)~dx$
from any binary string of length $n$.  A {\sl quadrature program\/} 
for $\int f(x)~dx$ is the mean of the estimates from several
primitive quadrature rules.

Section III develops
\begin{theorem}\label{theorem:NC}
Each Newton--Cotes estimate for
$\int f(x)~dx$ corresponds to a quadrature program based on
strings of low Kolmogorov complexity.
\end{theorem}

Basing these rules on computational complexity rather than
smoothness extends them 
to the case when there is no smoothness assumption on $f$.

Section IV discusses quadrature over higher dimensional domains.
The interpolation technique behind Newton--Cotes no longer
applies, but strings of low Kolmogorov complexity 
define new quadrature rules.  

Section V shows how concepts from 
Signal Reconstruction or Compressed Sensing (\cite{donoho})
provide information about the error terms, at least for ``sparse" functions.

Section VI speculates on further applications of these ideas.

\section{Quadrature Programs}

The Riemann integral $\int_0^1 f(x)~dx$ depends on having  
full information about $f$.  (By scaling and translation, 
restricting to integrals over the domain $[0,1]$ causes no
loss of generality.)  Briefly, 
the domain is subdivided, and $f$ is sampled in each subdomain.
One then takes the limit, as the mesh goes to zero, of the 
sums $\sum f(x_i^*) \Delta x_i$, where $x_i^*$ is the sample
point and $\Delta x_i$ is the size of the corresponding
subinterval.

Sampling and, therefore, computing the limit is not feasible when $f$ is known
by a table of values $f_i = f({i\over n})$, $i = 0,\ldots, n$.  In this case, 
one typically chooses one of the endpoints of each interval
as the sample point.

For convenience, let $h = 1/n$ denote the size of each subinterval.

\begin{definition}
A {\sl primitive quadrature rule\/} from the 
binary string {\tt b} for $f$ is the sum
${\hat f}_1h + \cdots + {\hat f}_nh$, where
\begin{displaymath}
{\hat f}_i = \Biggl \{
 \begin{array}{cc}
  f_{i-1} & {\rm\ if\ }{\tt b}_i = {\tt 0}\\
  f_i      & {\rm\ if\ }{\tt b}_i = {\tt 1}.\\
 \end{array}
\end{displaymath}
In other words, the binary string {\tt b} is an input to the pseudocode program
below.
\end{definition}

\medskip
\begin{flushleft}
{\tt              

float\ Quadrature(float\ f, bool\ b[],\ int\ n,\ float\ h)\ $\{$\\
\ \ int\ i\ =\ 0;\\
\ \ float\ q\ =\ 0.0;\\
\ \ for\ (i\ =\ 0;\ i\ <\ n;\ i++)\ $\{$\\
\ \ \ \ if\ (b[i]\ =\ 0)\ $\{$\\
\ \ \ \ \ \ q\ +=\ h*f[i];\ \ \ \ \ //\ left\ endpoint\\
\ \ \ \ $\}$\\
\ \ \ \ else\ $\{$\\
\ \ \ \ \ \ q\ +=\ h*f[i+1];\ \ \ //\ right\ endpoint\\
\ \ \ \ $\}$\\
\ \ $\}$\\
\ \ return\ q;\\
$\}$\\
} 
\end{flushleft}

\begin{definition}
A {\sl quadrature rule\/} is the estimate obtained from taking
the mean of the estimates from a finite set
of primitive quadrature rules.
\end{definition}

\subsection{Example}

Suppose that $n = 7$, so $f$ is defined by $f_0$, $\ldots$, $f_7$.  The
string {\bf b} $ = {\tt 0011101}$ yields the estimate

\begin{displaymath}
(f_0 + f_1 + f_3 + f_4 + f_ 5+ f_5 + f_7)h
\end{displaymath}
while {\bf b} $ = {\tt 0001111}$ yields

\begin{displaymath}
(f_0 + f_1 + f_2 + f_4 + f_5 + f_6 + f_7)h.
\end{displaymath}

\subsection{Strings of Low Complexity}

The binary strings of lowest complexity are {\tt 000$\cdots$0} and {\tt 111$\cdots$1}.
These correspond to using the left and the right endpoints of each interval,
respectively.  In the first case, though, the final value $f_n$ has no
effect on the estimate of the integral, while in the second 
the initial value $f_0$ is ignored.  A remedy to this situation is to take
the mean of the two estimates so obtained.  A simple calculation
shows that the estimate is then

\begin{displaymath}
\frac{f_0 + 2f_1 + \cdots + 2f_{n-1} + f_n}{2n},
\end{displaymath}
which is the well-known trapezoid rule.

\begin{proposition}
The trapezoid rule is the mean of the quadrature rules {\tt 0$\cdots$0}
and {\tt 1$\cdots$1}.
\end{proposition}

Alternatively, the trapezoid rule is the mean of the
quadrature rules {\tt 0101$\cdots$01} and {\tt 1010$\cdots$10}.

The next most complex strings are {\tt 0101$\cdots$01} and {\tt 1010$\cdots$10}.
{\sl Simpson's Rule\/} (\cite{burden}) estimates the integral as

\begin{displaymath}
\frac{f_0 + 4f_1 + 2f_2 + \cdots + 4f_{n-1} + f_n}{3n}
\end{displaymath}

\begin{proposition}
Simpson's Rule is the mean of the quadrature rules {\tt 0$\cdots$0}, {1$\cdots$1},
and {\tt 1010$\cdots$10}.
\end{proposition}

\section{Comparison with Newton--Cotes Quadrature}

More generally,
{\sl Newton--Cotes Integration\/} uses the Lagrange interpolation polynomial
of degree $n$ to derive an approximation that is exact when $f$
has degree $\le n$; see \cite{burden}.  Here are some common Newton--Coles
formulas, along with their interpretations as quadrature programs.  Notice that in
each case the complexity of the strings involved is quite low.
Also notice that in each case the estimate has the form $\sum_{i=0}^n a_i f_i$
with $\sum a_i = 1$.

\subsection{$n = 3$, or Simpson's Three--Eights Rule}

In this case, the function is sampled at four equally-spaced
points $(x_0, y_0)$, $(x_1, y_1)$, $(x_2, y_2)$, and 
$(x_3, y_3)$.

$$\int_{x_0}^{x_3} f(x)~dx \approx 
{{3h}\over{8}}\biggl [
y_0 + 3y_1 + 3y_2 + y_3
\biggr ]
.$$

This estimate is the mean of eight
primitive quadrature rules: three from {\tt 000}, three from {\tt 111}, 
one from {\tt 100}, and one from {\tt 110}.  To confirm, the {\tt 000} rule
yields $y_0 + y_1 + y_2$; the {\tt 111} rule yields
$y_1 + y_2 + y_3$; the {\tt 100} rule yields $y_1 + y_1 + y_2$;
and the {\tt 110} rule yields $y_1 + y_2 + y_2$.
These add up to $3 y_0 +  9 y_1 + 9 y_2 + 3 y_3$; divide by
8 to get the mean, and factor out the 3.

\subsection{$n = 4$}

In this case, the function is sampled at five equally-spaced
points $(x_0, y_0)$, $(x_1, y_1)$, $(x_2, y_2)$, 
$(x_3, y3)$, and $(x_4, y_4)$, and

$$\int_{x_0}^{x_4} f(x)~dx \approx 
{{2h}\over{45}}\biggl [
7y_0 + 32 y_1 + 12 y_2 + 32 y_3 + 7 y_4
\biggr ]
.$$

This corresponds to the mean of 45 primitive quadrature rules: 12 each from 
{\tt 0000} and {\tt 1111}, two from {\tt 0011}, and
19 instances from {\tt 1010}.

This is four of the samples of Simpson's Rule, plus seven more {\tt 1010}s
and two more {\tt 0011}.  The latter choice of endpoints 
concentrate on the center of the table, while the former concentrates
on the alternate endpoints.  

At this point the proof of Theorem~\ref{theorem:NC} is clear.

\section{Higher Dimensions}

The one-dimensional Newton-Cotes methods use an interpolating polynomial of degree
$d$.  One needs $d+1$ distinct
points to determine the coefficients of this polynomial.  This is easy
when the domain is an interval.

The situation is different in higher dimensions.  The dimension
of the vector space of polynomials
of degree at most $d$ in $n$ variables is
\begin{displaymath}
\biggl ( 
 \begin{array}{c}
   d+n \\ 
   n \\
 \end{array}
\biggr );
\end{displaymath}
this is the number of coefficients, or, since passing through a given point 
imposes one linear constraint on the polynomial, the number of points
required to determine the coefficients uniquely.

The number of points in a cubic grid is $2^d$, but adjoining adjacent cubes 
leads to other grid point counts.  The difficulty is matching the number of
grid points to the number of coefficients.  As a rule, this is impossible.

Any sequences of digits modulo $2^n - 1$ still determines a
primitive quadrature rule.
Look, for example, at an $m\times m$ array in dimension 2, which is made
up from $m^2$ {\sl primitive\/} ({\it ie\/}, $2\times 2$) squares.  Arbitrarily label the corners of each square 0, 1, 2, and 3, for example starting at the northwest corner 
and proceeding clockwise.

Now, consider the mean of the four low--complexity sequences
{\tt 000$\dots$0}, {\tt 111$\ldots$}, {\tt 222$\ldots$}, and {\tt 333$\ldots$}.
(Each has length $m^2$.)
In each primitive square, the corresponding entry in the sequence determines which
grid point to choose.

The result is a quadrature rule that weights each of the corner points with weight
1, each of the non--corner edge points with weight 2, and each of the 
interior points with weight 4; the weighted sum is then divided by 4.  

\begin{theorem}
The mean of the four low--complexity sequences
{\tt 000$\dots$0}, {\tt 111$\ldots$}, {\tt 222$\ldots$}, and {\tt 333$\ldots$}
defines a quadrature rule with weights

\smallskip
\begin{tabular}{cccccc}
&1        & 2 & $\cdots$ & 2 & 1\\
&2        & 4 & $\cdots$ & 4 & 2\\
&$\vdots$ & $\vdots$ & $\cdots$ &$\vdots$ & $\vdots$\\
&2        & 4 & $\cdots$ & 4 & 2\\
&1        & 2 & $\cdots$ & 2 & 1.\\
\end{tabular}
\IEEEQED
\end{theorem}

\section{Error Results}


So far, there has been no mention of an error estimate of the kind
associated with the   Newton--Cotes rules.  
Notice 
that all of the Newton--Cotes rules take as input the same table of values $\{f_i\}$, but
higher order estimates give a tighter error bound.  While there
is no more information to be gained from the the table of values,
the extra smoothness assumptions of the higher--order methods
seem to provide more information about the function itself. 

The error of one of these estimates is zero when the sampled
function is a polynomial of sufficiently low degree.  When the function
is a polynomial of low degree then the table of values can
contain no more information than the ordered set of coefficients.
The heuristic is that {\sl more\/} information {\sl about a function
enables a tighter error estimate.\/}

Recently, Donoho (\cite{donoho}) and others have
investigated the problem of Compressed Sensing (CS), which is to reconstruct a 
signal represented as a vector from a sample of its entries.  
Donoho  showed that
knowing that a vector can be compressed is enough to reconstruct
it, even without knowing what the compressed version might have been.
When integrating the goal is to {\sl process\/} the signal rather
than to reconstruct it, but the same principle applies.  

This section contains two results.  The first, following 
Donoho (\cite{donoho}), relates the information in the error in 
an integral estimate to the error in the information in the 
description of the integrand, a kind of 
{\sl Error--Information Duality\/}.  The second proves 
that for sparse functions (see below) there exists a quadrature
program estimating the integrl to arbitrary precision.

\subsection{Error--Information Duality}

Following Donoho, the functions of interest have the form 
$f(x) = \sum a_j \phi_j(x)$, where the functions $\phi_j$ 
form a basis for an appropriate space of functions.  
(The space for which they form a basis is intentionally left 
vague in order to be as general as possible.)
The function is {\sl sparse\/} if for some $R > 0$, 
$$\Vert a \Vert_p < R,$$
where $0 < p < 2$ and $\Vert {\bf a} \Vert_p$ is the $l^p$
norm of the series of coefficients $a_1, a_2, \ldots$.  A 
function whose expansion has many small terms fails to be
sparse by this definition, while a finite degree polynomial expansion
is sparse.

Let $X_{p,n}(R)$ denote the space of functions given by a table
of $n$ values which are $l^p$ sparse in the sense above. 
This is the space of functions of interest. 

Begin with the functions $f$ with  $d+1$ nonzero coefficients, generalizing
the space of polynomials if degree $\le d$.  Renumber
if necessary so that the nonzero coefficients are $a_0,
\ldots, a_d$.

The entries in the table of values ${\bf f} = [ f_0 
\cdots f_n]^T$ are  
$\sum _{j=0}^d a_j \phi_j({i\over n})$.
Let $\Phi$ denote the $(n+1)\times d$ matrix with
entries $\phi_j({i\over n})$.  Let ${\bf a} = 
[ a_0 \cdots, a_d]^T$.  Then ${\bf f}
= \Phi {\bf a}$.  The matrix $\Phi$ only depends on
the basis $\{\phi_j\}$.

Now, integrate $f$.  First,  let $q_j = \int_0^1 f_j(x)~dx$,
and let ${\bf Q} = [ q_0, \ldots, q_d ]^T$; like
$\Phi$, {\bf Q} only depend on the basis.  Since $f(x) = \sum_0^d a_j \phi_j(x)$, 
 $\int_0^1 f(x)~dx = \sum_0^d a_j q_j$
 $= {\bf Q}{\bf a}$.

Next, suppose that $\Phi$ has a left inverse $\Phi^{-1}$, noting that
this is never the case when $n+1 < d$.   
Then ${\bf a} = \Phi^{-1}{\bf f}$, and

\begin{theorem}
When the expansion of $f$ has $d+1$ coefficients and $n > d+1$
then
$\int_0^1 f(x)~dx = {\bf Q}\Phi^{-1}{\bf f}.$
\IEEEQED
\end{theorem}

Compare this theorem with the exactness results for 
Newton--Cotes integrals of degree $d$ polynomials.

When there are more than $n$ nonzero coefficients, the integral
can be estimated by truncating the series expansion to include
the $n$ ``most important" coefficients.  The truncated function
is integrated exactly, so the error in the estimate comes from
the coefficients that were ignored.  The truncated function 
contains $n$ {\tt float}s worth of information, plus a little more 
to describe where these coefficients are in the series expansion.
The table of values has $n$ {\tt float}s worth of information
as well.  The information in the error in the integral estimate
is exactly the information in the ignored coefficients.  Hence

\begin{theorem}[Error--Information Duality]
The information content of the error is 
(a digest of) the error in the 
known information about the integrand.
\IEEEQED
\end{theorem}

\subsection{Good Quadrature Rules}

Now suppose that $f$ is a sparse function in the sense of the section
above, so that there exists a good estimate
\begin{displaymath}
\leqno{(\dagger)}\int_0^1f(x)~dx = \sum_{i=0}^n a_if_i.
\end{displaymath}

This section shows

\begin{theorem}
For any $\varepsilon > 0$ there exists a quadrature program 
that approximates ($\dagger$) within $\varepsilon$.
\end{theorem}

{\sl Proof.\/} Choose rational numbers $y_i/r$ such that sup$\{|a_i - y_i/r|\}
< \varepsilon/n.$  Here $r$ is any convenient common denominator.  Notice
that $\sum y_i = r$, because of the weighted average nature of $(\dagger)$.
The proof finds quadrature programs that reproduce the coefficients
$y_i/r$.

Choose $r$ quadrature rules $\QR 1 l, \QR2 l, \ldots, 
\QR n l$ where $l$ runs from 1 to $r$.  Each $\QR i l$ leads to
an estimate as in Section II, part A.  

Now, consider the contribution of each $f_i$.  The only contribution
from $f_0$ occurs when $\QR 1 l = 0$, so 

\begin{displaymath}
y_0 = h\sum_{l=1}^r (1 - \QR 1 l) ,
\end{displaymath}

The only contribution from $f_n$ occurs when $\QR n l = 1$, so

\begin{displaymath}
y_n = h\sum_{l=1}^r \QR n l .
\end{displaymath}

The contribution from $f_i$, where $i$ is neither $1$ nor $n$ occurs
when $\QR {i - 1} l = 0$ (left endpoint) or when $\QR i l = 1$, so

\begin{displaymath}
y_i = h\sum_{l=1}^r (1 - \QR {i-1} l +\QR i l) .
\end{displaymath}

Next, solve for the $\QR i l$.  From the $f_0$ coefficient, h$\sum \QR 0 l = 
r - y_0$.  Plug this into the relation for the $f_1$ coefficient, so
$y_1 = h\sum_{l=1}^r (1 - \QR 0 l +\QR 1 l)$, implying that
$h\sum \QR 1 l = 2r - y_0 - y_1$.  Continuing in this way
shows that $h\sum \QR i l = ir - y_0 - y_1 - \cdots - y_i$

Finally, $y_n = h\sum \QR n l$, but this is redundant since the
$f_{n-1}$ coefficent satisfies $h\sum \QR {n-1} l = nr - y_0 - y_1 - \cdots - y_{n-1}$
Since $\sum y_i = 1$, the theorem is proved. \IEEEQED

\section{Further Work}

One foresees two kinds of further work.  The first involves
the concept of integration.  Suppose that one makes a random
choice of binary string(s) to define a quadrature program: what is the
probability that this program is good?  The sample space here is
well--defined, namely, binary strings, but the concept of ``good" 
needs refinement, especially with regard to the space of functions
to be integrated.  Integrating smooth functions allows one to compare
the results with Newton--Cotes quadrature, but seems excessively
restrictive in terms of the applications in the introduction.
Perhaps it would be better to survey, say, $L^2$ functions,
by choosing random coefficient for a wavelet \cite{daubechies} basis.

There is also further work possible from the perspectives
of signal processing, compressed sensing, and cryptography.
One way to think of $\int_0^1 f(x)~dx$ is to think of $f$ as
a message and the integral as a {\sl message digest\/}.  From
a cryptographic perspective, this is not a good message digest, because
the information from the high order bits of the message has no
effect on the low-order bits of the digest, while an ideal message digest
should appear random.  Can one characterize
other message digests in terms the information content added
by the algorithm?  Is this a measure of security?

From the signal processing perspective, the function $f$ represents
some signal and the integral is a simple form of  on-line processing.  
It is a simple matter to integrate against a kernel $K(t)$, that is,
to estimate $\int K(t) f(t) ~dt$, as long as onehas enough information
about $K$.  But what of more complex
processes like convolution?  These problems are particularly
interesting in the context of compressed sensing: what is
the information-theoretic meaning of an integral transform
when the function $f$ is compressible?

\section*{Acknowledgment}

The author thanks Leonid Hanin, Larry Kratz, and
William McCurdy for stimulating conversations on this subject.

\ifCLASSOPTIONcaptionsoff
  \newpage
\fi



\begin{thebibliography}{1}

\bibitem{burden}
R.~L.~Burden and J.~D.~Faires, {\sl Numerical Analysis\/}.
Brookes--Cole (2004).

\bibitem{daubechies} I.~Daubechies, {\sl Ten Lectures on
Wavelets\/}.  NY: SIAM (1992).

\bibitem{donoho} D.~L.~Donoho, ``Compressed Sensing."
{\sl IEEE Transactions on Information Theory\/} vol.~52, no.~4, April,
2006.

\bibitem{hamming}
R.~W.~Hamming, {\sl Numerical Methods for Scientists and
Engineers\/}, Second Edition.  NY: Dover(1986).

\bibitem{shannon}
C.~Shannon, ``A Mathematical Theory of Communication,'' 
{\sl Bell System Technical Journal\/} vol.~27 (1948), 
623--56.

\end{thebibliography}
\end{document}